\documentclass[aps,pra,twocolumn,floatfix,amssymb]{revtex4-2}

\usepackage[scr=boondox]{mathalpha}

\usepackage{graphicx,dcolumn,bm,xstring,amsmath,placeins}
\usepackage{color}
\usepackage[colorlinks,allcolors=blue,bookmarks=true]{hyperref}

\graphicspath{{./},{./Figs/}} 

\newcommand{\mass}{\mathsf{m}}

\newcommand{\ket}[1]{\left| #1 \right\rangle}
\newcommand{\braket}[1]{\left\langle #1 \right\rangle }

\newcommand{\Braket}[2]{\left\langle #1 \middle| #2 \right\rangle}

\newcommand{\tr}{\mbox{\text{Tr}}}

\newcommand{\beq}{\begin{eqnarray}}
\newcommand{\eeq}{\end{eqnarray}} 

\newcommand{\hide}[1]{}  

\newcommand{\TODO}[1]{} 
 
\newcommand{\Eq}[1]{{\textcolor{blue}{Eq.}}~\!\!(\ref{#1})} 
\newcommand{\Sec}[1]{{\textcolor{blue}{Sec.}}~(\ref{#1})} 
\newcommand{\App}[1]{{\textcolor{blue}{Appendix}}~\ref{#1}} 
\newcommand{\Fig}[1] {{\textcolor{blue}{Fig.}}~\!\!\ref{#1}}
\newcommand{\sect}[1]{\paragraph{#1.--}}

\newcommand{\hrefl}[2]{\href{#2}{(#1)}}

\begin{document}
\title{Mesoscopic superfluid to superconductor transition} 

\author{Yehoshua Winsten, Doron Cohen} 

\affiliation{
\mbox{Department of Physics, Ben-Gurion University of the Negev, Beer-Sheva 84105, Israel} 
}

\begin{abstract}
Spectrum tomography for the energy~($E$) of a ring-shaped Bose-Hubbard circuit is illustrated. There is an inter-particle interaction~$U$ that controls superfluidity (SF) and the transition to the Mott Insulator (MI) regime. The circuit is coupled to an electromagnetic cavity mode of frequency~$\omega_0$, and the coupling is characterized by a generalized fine-structure-constant $\alpha$ that controls the emergence of superconductivity (SC). The ${(U,\alpha,\omega_0,E)}$ diagram features SF and SC regions, a vast region of fragmented possibly chaotic states, and an MI regime for large~$U$. The mesoscopic version of the Meissner effect and the Anderson-Higgs mechanism are discussed. 
\end{abstract}

\maketitle


\section{Introduction} 

The superfluid (SF) to Mott insulator (MI) transition in the Bose-Hubbard (BH) model stands out as a paradigmatic example for quantum phase transitions \cite{Leggett,BHHqpt,BHH1,BHH2}. It provides a simplified description of interacting bosonic particles confined in a lattice potential. In the SF regime one observes that the system can have both meta-stable condensates and fragmented (FR) states. The former support SF, while the latter are accessible once the system thermalizes.   
%
Closely related is superconductivity~(SC) \cite{Leggett2,Leggett3}. The theory of SC has two faces: On the one hand we have the BCS theory that clarifies why electrons behave in some sense as a bosons, due to Cooper pairing. The other aspect is the Meissner effect, which is related to the interaction with the electromagnetic (EM) field, and to the Anderson-Higgs mechanism \cite{PWA1,PWA2,PWA3,TDGL}. It is the latter aspect that makes the notion of SC distinct from SF.          

\sect{Minimal model for SF}
The BH model serves as a minimal model for the explanation of SF. Namely, the superflow becomes metastable thanks to the inter-particle interactions. But if the interaction is further increased, the ground state changes from coherent condensation, to a fragmented site-occupation, which is the SF-MI transition \cite{MISF1,MISF2,MISF3}. Formally speaking, this transition is abrupt only for an infinite chain, but its mesoscopic version is clearly apparent even for the two-site model, as pointed out long ago by Leggett \cite{Leggett}, who called it a transition from Josephson-regime to Fock-regime \cite{csd,DimerQPT}. For a ring with several sites, condensation in an excited momentum orbital implies the feasibility of superflow. Additionally, one has to address the significance of the underlying {\em chaos}. Recent studies of the chaos-mediated SF-MI transition are \cite{BHHchainChaos1,BHHchainChaos2} and \cite{bem}. The latter adopts a {\em tomographic approach} that we would like to pursue.

\sect{Minimal model for SC}
The question arises, what is the minimal model for the discussion of the SF-SC-MI transition and the Meissner effect. We take here literally the old idea that SC can be regarded as arising from condensation of charged bosons. Accordingly, the minimal model consists of a BH ring coupled to a single EM mode of a cavity. The latter is modeled as an LC circuit.  This configuration is enough to demonstrate the emergence of SC.   
The dimensionless parameter that characterizes the coupling between the BH ring the the EM mode is a dimensional fine structure constant $\alpha$. This parameter controls the crossover from the SF regime to an SC regime. 
The excitation frequencies of the system are affected by this EM interaction, which constitutes a mesoscopic version of the Anderson-Higgs mechanism. 
On top we have the inter-particle interaction that can induce a transition to an MI regime.

\sect{Quantum chaos}
The SF-MI transition is apparent also in the parametric evolution of the {\em excited states}.  In this context, quantum chaos is an aspect that should not be overlooked \cite{dys2,kolovskiPRL,trimerKottos,BHHchains-Henn,MixedChaos,sfc,sfa,BHHtrimerChaos,BHHchainChaos1,BHHchainChaos2}. 
The common ``quantum chaos" practice to characterize a system is to look on its energy level statistics. But such approach has two issues: {\bf (i)}~It provides numerically clear results only for rather simple systems with large number ($N$) of particles, such that the spectrum in the range of interest is dense enough; {\bf (ii)}~It does not allow classification of the eigenstates in the typical situation of underlying mixed chaotic and quasi-regular dynamics. In the present context, one would like to highlight meta-stable SF states that are embedded in a background of FR states.  

\sect{Tomography}
As opposed to the common practice, our approach is {\em numerically cheap} and flexible enough to deal with small complex systems. We use the term {\em quantum phase-space tomography} \cite{bem} in order to emphasize that the inspection of the spectrum is not limited to ``level statistics", neither focused on the ground state, and possibly reveals the relation to the underlying phase space structure. The latter serves as a classical skeleton for the quantum eigenstates.

\sect{Outline}
The circuit Hamiltonian and its parameters are introduced in \Sec{sModel} and \Sec{sModelP}.  
The big picture regarding SF/SC superflow is provided in \Sec{sSuper}. 
The topography of the energy landscape is discussed in \Sec{sFloor}, with focus on the Landau perspective. 
The semiclassical SF-SC border is explained in \Sec{sBorder}. 
Tomography measures are defined in \Sec{sMeasures} and applied in \Sec{sTomog}. 
The various regions in the quantum spectrum are further elaborated in \Sec{sQborders}. 
The mesoscopic Meissener effect and its relation to the Anderson-Higgs mechanism are discussed in \Sec{sMeiss}.  
We finalize with a summary in \Sec{sF}. 
The extra appendices \App{sCavity} and \App{sSQUID} and \App{sLondon} help to place this work in the common context of SC studies.

\clearpage
\section{The model} 
\label{sModel}

We consider a circuit that is modeled as a Bose-Hubbard ring that interacts with a cavity mode. The EM mode is described by conjugate variables ${(Q,\Phi)}$ that essentially encodes the electric field and the magnetic flux, see \App{sCavity}. The dynamics of the bosons on the ring is generated by operators $\bm{a}_j$ and $\bm{a}_j^{\dag}$ that annihilated and create particles in sites that are indexed by~$j$. Additionally,     
in Coulomb gauge, the Hamiltonian contains also a term $\mathcal{U}_{ES}$ for the extra electrostatic energy.
\beq \label{eHraw}
\mathcal{H} = 
&&  \frac{U}{2}\sum_j \bm{a}_j^{\dag}\bm{a}_j^{\dag}\bm{a}_j\bm{a}_j   
-\frac{K}{2}\sum_{j=1}^{L}  \left[ \bm{a}_{j{+}1}^{\dag} \bm{a}_j \ e^{i (1/L) e\Phi} 
+ \text{hc}\right]    
\nonumber \\
&& \ + \ \mathcal{U}_{ES} \ + \ \frac{c^2}{2L_e}\Phi^2   
\ + \ \frac{1}{2C_e} Q^2
\eeq 
where $L_e$ and $C_e$ have dimension of length,    
and reflect the geometry of the system. 
In particular, the bare frequency of the EM-mode 
reflects the linear size $\mathcal{R}$ of the cavity 
\beq \label{eOmega}
\omega_0 \ = \ \frac{c}{\sqrt{L_eC_e}} \ \equiv \ \frac{c}{\mathcal{R}}
\eeq  
As explained in \App{sCavity}, 
typically ${L_e \sim \mathcal{L}^2/\mathcal{R} }$, 
where $\mathcal{L}$ is the linear size of the ring. 
We choose convenient units of charge 
\beq
q^2 \ \ = \ \ \sqrt{\frac{C_e}{L_e}} c 
\eeq
and define 
\beq
\bm{b} \ \ = \ \ \frac{1}{\sqrt{2}} \left[ q\Phi + \frac{i}{q} Q \right]
\eeq
The dimensionless flux is 
\beq
\bm{\phi} \ \ = \ \ e\Phi \ \ = \ \ \sqrt{\frac{\alpha}{2}} (\bm{b}^{\dag}+\bm{b})
\eeq
where the generalized fine structure constant is 
\beq \label{eAlpha}
\alpha \ \equiv \  \frac{e^2}{q^2} \ \sim \ \frac{e^2}{c} \left( \frac{\mathcal{L}}{\mathcal{R}} \right)^2
\eeq
The Hamiltonian takes the form
\beq \label{eH}
\mathcal{H} = 
&& \ \ \omega_0 \bm{b}^{\dag}\bm{b} 
+ \frac{U}{2}\sum_j \bm{a}_j^{\dag}\bm{a}_j^{\dag}\bm{a}_j\bm{a}_j  
+ \mathcal{U}_{ES}
\nonumber \\
&& - \frac{K}{2}\sum_{j=1}^L  \left[ \bm{a}_{j{+}1}^{\dag} \bm{a}_j \ e^{i (1/L) \sqrt{\frac{\alpha}{2}} (\bm{b}^{\dag}+\bm{b})} + \text{hc}\right]  
\eeq
The explicit expression for the extra electrostatic energy $\mathcal{U}_{ES}$ of \Eq{uES} as a function of the occupation operators $\bm{a}_j^{\dag}\bm{a}_j$ will be provided in a later section. 

We switch to the momentum representation: 
\beq 
\mathcal{H} = 
&& \frac{\omega_0}{2} \left[\frac{\alpha}{2} \bm{n}_{osc}^2 + \frac{2}{\alpha} \bm{\phi}^2\right]
- K \sum_{k} \bm{n}_k \cos\left( k - \frac{\bm{\phi}}{L} \right) 
\nonumber \\ \label{eHn} 
&& \ + \mathcal{U}_{ES} \ -\frac{U}{2L} \sum_k \bm{n}_k^2  \ + \text{transitions} 
\eeq 
where ${\bm{n}_{osc}=(1/e)Q}$, and $\bm{n}_k$ are the occupations of the momentum orbitals. The latter are indexed by 
\beq
k = k_m = \frac{2\pi}{L} m,
\ \ \ \ \ \ \
\text{[$m=$ integer mod($L$)]}   
\eeq
The trimer ($L{=}3$) has three orbitals that we label as $k_0$ and $k_{\pm}$. 
In the numerical code the standard basis consists of the occupation states ${\ket{n_k, n_{osc}}}$. These are the eigenstates of the $U{=}0$ Hamiltonian. Condensation in the $k_m$ orbital means ${n_k=N\delta_{k,k_m}}$.

\section{Model parameters} 
\label{sModelP}

The parameters ${(L, \ N, \ N_{osc}, \ K, \ U, \ \omega_0, \ \alpha)}$ define the Hamiltonian \Eq{eH}, where $N_{osc}$ is an insignificant arbitrary numerical truncation of the oscillator's Hilbert space for numerical purpose.  
We want to define set of dimensionless parameters that appear in the {\em classical} equations of motion. For the ring the scaled coordinates are $\tilde{\bm{a}}_j=\bm{a}_j/\sqrt{N}$, while for the oscillator we keep $\phi$ and define consistently ${\tilde{\bm{q}}={\bm{q}}/N}$, such that ${\hbar_{scaled}=1/N}$ for all the canonical coordinates. Consequently, the implied classical Hamiltonian is ${H_{cl}=\mathcal{H}/N}$. Without loss of generality we choose the units of time such that $K{=}1$. Accordingly the {\em classical} dimensionless parameters are 
\beq \label{ePar}
\left(u \equiv \frac{NU}{K}, \ \ \omega_0/K, \ \ N\alpha\right)
\eeq
Additionally, we have to specify the number of sites $L$.
Assuming a large multi-site ring ${L \gg 1}$, 
with lattice spacing $\ell$, 
We can define effective Gross-Pitaevskii Equation (GPE) parameters, namely, 
the mass ${M=1/(K\ell^2)}$, 
the interaction ${g=\ell U}$, 
and the length ${\mathcal{L}= L\ell}$. 
It is convenient to define dimensionless parameters that survive the continuum limit, such that the dependence on $\ell$ cancels out. These are 
\beq \label{euL}  
u_L &=& 
L\frac{NU}{K} 
=  \mathcal{L} N M g  
\equiv \left(\frac{\mathcal{L}}{\xi}\right)^2 
\\ \label{ewL} 
w_L &=& 
\frac{1}{L^2} \frac{N\alpha}{\omega_0/K} 
= \frac{1}{\mathcal{L}^2} \frac{N \alpha}{\mass \omega_0} 
\equiv \left(\frac{\mathcal{R}}{\lambda}\right)^2
\eeq
Later we discuss their significance, and clarify that $u_L$ and $w_L$ are the {\em classical} dimensionless parameters that control the emergence of SF and SC respectively. 

In the quantum treatment we have to specify also~$N$. Note again that in the so-called second quantization scheme $1/N$ plays the role of a dimensionless Planck constant. As explained in \cite{bem} (and references therein) the Mott transition is controlled by the {\em quantum} dimensionless parameter 
\beq \label{eGamma}
\gamma_L = \frac{u_L}{N^2}
\eeq  
In the continuum GPE limit this parameter controls the crossover to the hard-core bosons regime where the particles can be regarded formally as fermions.  

The Hamiltonian \Eq{eH} is a closely  related to the familiar model of coupled Josephson junctions. Assuming equal site occupation (${n_j:=N/L}$), it is like the SQUID Hamiltonian of \App{sSQUID} with ${E_C=U}$ and ${E_J=NK/L}$. Thus we make the identification ${\gamma_L = E_C/E_J}$, and point out that $u$ would become meaningless (for example, there is no meaning to discuss $u{=}0$ condensation in momentum orbitals).           

The parameter $\xi$ that is defined via \Eq{euL} is the standard definition of the healing length in the GPE context. The parameter $\lambda$ that is defined via \Eq{ewL} is motivated by the discussion of the London penetration depth in \App{sLondon}. For a thin ring it is merely a formal notation, and the ratio ${\kappa=\lambda/\xi}$ has no practical meaning.

\section{Superflow}
\label{sSuper}

The essence of superfluidity (SF) and superconductivity (SC) is the feasibility of superflow. This superflow is a metastable condensation of the particles in an excited momentum orbital. As opposed to that, we have eigenstates with fragmented (FR) occupation that possibly support microscopic rather than macroscopic persistent current. 
In this section we summarize the ``big picture" regarding the emergence of superflow and its dependence on the major dimensionless parameter of the model. In the subsequent sections we provide a more detailed account.    

\subsection{Superfluidity (SF)}

If ${\alpha=0}$, the variable $\phi$ becomes constant of motion, and the Hamiltonian of the ring becomes formally identical to that of an SF circuit in a rotating frame, where $\phi$ is the Sagnac phase (proportional to the rotation velocity). Depending on the strength of the interaction, such ring can support superflow: a metastable persistent current due to condensation of particles in an excited momentum orbital. 

In the absence of interaction ($u=0$), the ground-state of a non-rotating ring ($\phi=0$) is the condensation in the ${k_0=0}$ orbital. As we turn on the interaction (keeping $\phi=0$), additional condensates at ${k_m}$ may become metastable. We refer to this as SF. To be precise, due to the interaction the ground-state and the metastable states become {\em squeezed} coherent states, meaning that the occupation of the $k_m$ orbital exhibits some depletion. Unlike the ${m=0}$ ground-state, the ${m\ne0}$ metastable states are immersed in quasi-continuum of FR states that are possibly supported by chaotic regions in phase-space~\cite{bem}. Quantum mechanically their meta-stability is endangered by tunneling.  

An equivalent way to describe the meta-stability is to say that the ${k_0}$ condensate remains metastable in a rotating frame, namely, within a range ${|\phi|<\phi_c}$, where $\phi_c$ is the (dimensionless) critical velocity that is implied by the Landau criterion, which we further discuss below in subsection~C of \Sec{sFloor}.

\subsection{Superconductivity (SC)}

The other way to gain metastability (not related to~$U$) is encountered once $\phi$ becomes a dynamical coordinate. We refer to this as SC. This coordinate can adjusts itself such that the $k_m$ condensate becomes meta-stable. The idea is easily understood if one assumes ${u=0}$. Then the occupations of the orbitals become constant of motion. Dropping a constant, we get for a given occupation ${\bm{n}_k \mapsto n_k}$ the following Hamiltonian 
\beq \label{eHosc}
\mathcal{H}_{osc} =  
\frac{\omega_0}{2} \left[\frac{\alpha}{2} \bm{n}_{osc}^2 + \frac{2}{\alpha} \bm{\phi}^2\right]
- N_sK \cos\left(\frac{\bm{\phi}}{L}-\bar{k}\right)
\ \ \ \ \  
\eeq 
In the above expression the definitions of $N_s$ and $\bar{k}$ are implied via \Eq{eHn}. For condensation in the $k_m$ orbital we get ${\bar{k}=k_m}$, and $N_s=N$ attains its maximal value. If $\alpha$ is large enough, $\phi$ can adjust to a value ${\phi \sim \phi_m}$, where 
\beq
\phi_m = 2\pi m, \ \ \ \ \ [m=\text{integer}] 
\eeq
such that the condensate becomes metastable. The energetic price for that is ${E \sim (\omega_0/\alpha)\phi_m^2}$. Due to this symmetry breaking we get an SC-type superflow. Irrespective of that, most eigenstates feature fragmented (FR) occupation for which the effective $N_s$ is smaller compared with~$N$.    

For non-zero interaction~$u$ \Eq{eHosc} can be regarded as an effective approximation. The condensates are no longer exact eigenstates. Due to the squeezing they exhibits some depletion, which implies that they do not feature maximal $N_s$. For very large $u$ the ground state becomes MI, with equal occupation of the orbitals, for which ${N_s=0}$.

\section{The energy landscape}
\label{sFloor}

The energy landscape is dictated by 3~energy scales
\beq
W_{K} &=& \frac{NK}{L^2}  \\
W_{U} &=& \frac{N^2U}{L}  \\
W_{\Phi} &=& \frac{\omega_0}{\alpha}  
\eeq 
Note that division of the $W$-s by $N$ are the ``classical" parameters \Eq{ePar} of the model.  Their significance is as follows: Up to numerical prefactor, $W_{K}$ is the price of having a condensation in the next excited orbital ${k_1=2\pi/L}$. Looking on the 3rd term of \Eq{eHn} one realizes that due to the interaction a condensate has lower energy compared with a fragmented state, where the price of fragmentation is $W_{U}$. Finally, in the SC context, $W_{\Phi}$ reflects the price of non-zero flux ${\phi_1=2\pi}$. 
The associated dimensionless parameters $u_L$ and $w_L$ of \Eq{euL} and \Eq{ewL} 
are the ratios ${W_U/W_K}$ and ${W_K/W_{\Phi}}$ respectively.  
If $u_L$ is too large, the condensate get fragmented, aka Mott transition, and the superflow is diminished. The dimensionless parameter that controls this transition is ${\gamma_L}$ of \Eq{eGamma}. Note that in the SF context finite $u$ is essential to create the metastability, but if $u$ is too large, it diminishes this metastability. As opposed to that, $u$ is not required for getting the SC metastability.

\begin{figure}

\includegraphics[width=5cm]{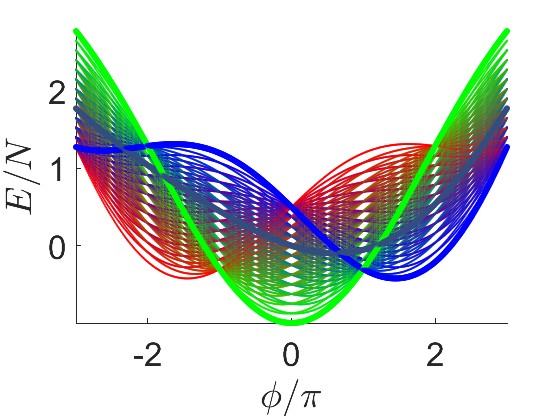}
\includegraphics[width=5cm]{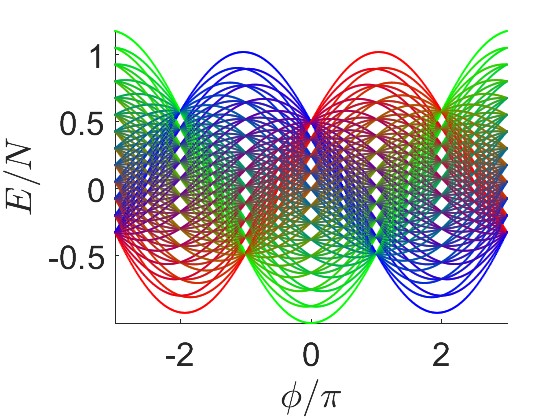}
\includegraphics[width=5cm]{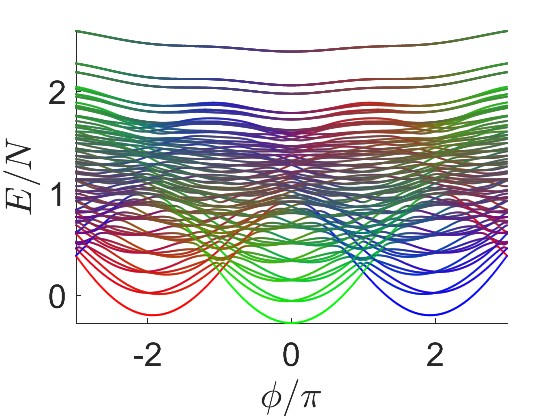}

\caption{
{\bf Born-Oppenheimer energies.}
The 91 energy levels of an  $N=12$ trimer as function of $\phi$. From up to down, the model parameters are ${u=0,0,5}$ and ${\alpha=2,20,20}$, while $\omega_0=0.48$. The units of time here and in subsequent figures are chosen such that ${K=1}$. We use RGB color-code for the occupation as explained in the text. 
Most of the levels cannot be resolved, and therefore in the left panel we highlight representative energy levels: full condensation in the $k_0$ orbital (green); full condensation in the $k_{+}$ orbital (blue); and fragmented 20\% 30\% 50\% occupation (grayish).
}  
\label{fBO}  
\end{figure}

\subsection{Potential floor}

In order to figure out what are the metastable flow states of the system, it is convenient to regard ${\ket{\phi, n_k}}$ as the configuration space. Then one can define the potential floor as 
\beq
V(\phi, n_k)  \ = \ \min[\mathcal{H}]
\eeq
where the minimum is taken with respect to the conjugate coordinates. Note that at this stage of the analysis we adopt a semiclassical perspective and ignore quantum uncertainties.    

In order to get visualization of the energy landscape let us first neglect in \Eq{eHn} the interaction induced transitions, meaning that the occupations $n_k$ of the orbitals become good quantum numbers. Each $\{ n_k \}$ configuration defines via \Eq{eHosc} a potential-energy that can be regarded as the {\em Born-Oppenheimer} level with regard to the $\phi$ coordinate. We refer to them below as {\em washboard-sections}. The major washboard-sections correspond to $k_m$ condensates defined as ${n_k=N\delta_{k,k_m}}$. The intermediate washboard-sections correspond to fragmented occupation. 

{The Born-Oppenheimer washboard sections are illustrated in \Fig{fBO} for an $L=3$ ring. The RGB color-code reflects the occupation such that red, green and blue indicate condensation in one of the 3~orbitals.}

\subsection{Valley structure}

The potential floor feature a {\em valley structure}. There is a major valley around the $k_0$ condensate at $\phi_0$, and possibly there are several metastable valleys around the $k_m$ condensates at ${\phi \sim \phi_m }$. 
In the illustration of \Fig{fBO}a the lowest valley is formed by the green colored sections, and the two metastable valleys are formed by the red and blue colored sections. 
The energies of the meta-stable minima are ${E \sim W_{\Phi} m^2}$, and they are separated by barriers whose height is of order~$W_K$. 
Accordingly, the energy floor features $\sim w_L$ major valleys. 
For large enough $\alpha$ there are $L$ major valleys, each correspond to a different condensate. They are located at the smallest values of ${\phi_m}$. Due to the discreteness of the ring there might be also secondary valleys, for larger ${|m|}$ values. 

{In the numerical demonstration we forcus on the 3 major vallies of an $L=3$ ring. If we used larger $L$, there would be more `color-distinct' major valleys. Other than that, as long as the `transitions' in \Eq{eHn} are neglected, the overall topography of potential floor depends on~$L$ merely via the classical dimensionless parameters. If we take the `transitions' into account, then $L$ has a secondary effect on the appearance of the Landau {\em grooves}, as further discussed below. These grooves are important for the understanding of SF metastability, but less important for the discussion of SC.}        

In the absence of interaction each whasboard-section holds eigenstates that can be labeled by the occupations $n_k$ and by a running index ${\nu_{osc}=0,1,...}$. Those unperturbed eigenstates can be categorized into two groups: those that reside inside the valleys, and those that reside ``above" the energy barriers. As the interaction is switched on, states of different washboard-sections are mixed, unless they reside inside valleys that are separated by an energy-barrier.

\subsection{Landau Grooves}
\label{sGrooves}

The major washboard-sections that correspond to $k_m$ condensates are separated by intermediate washboard-sections that correspond to fragmented occupation. The 3rd term of \Eq{eHn} implies that due to an energy cost $W_U$, a condensates might have lower energy compared with fragmented states of the nearby washboard-sections. To be more precise, the valley that is associated with the $k_m$ condensate is centered at $\phi_m$, and its stretch along the $\phi$ direction is determined by the Landau criterion as explained below. The stretch looks like a {\em groove} along the potential floor ${V(n_k,\phi)}$.    
       
Holding $\phi$ fixed is formally like considering a rotating ring. A condensate, say at $k_0$, is a local minimum of the potential if all the Bogolyubov frequencies are real and positive. The lowest excitation frequency $\omega_q$ is given by Eq(B2) of \cite{sfa}) with ${q=\pm 2\pi/L}$.
If the ring is large (${L\gg 1}$) an approximation is 
${\omega_q \approx c |q| \pm v_{\phi}q }$, 
where $c =(K/L)\sqrt{u_L}$ is the speed of sound 
and $v_{\phi} \approx (K/L)\phi$ is the rotation velocity of the ring.
Hence we get the Landau criterion for stability ${|\phi|<\phi_c}$, 
where  ${\phi_c=\sqrt{u_L}}$. More generally, if $L$ is not large, 
it is straightforward to deduce that critical value is determined by    
\beq
2 \cos\left(\frac{\phi_c}{L}\right) = -\left(\frac{u_L}{L^2}\right) 
+ \sqrt{ \left(\frac{u_L}{L^2}\right)^2 + 4 \cos^2\left(\frac{q}{2}\right)} 
\ \ \ \ \ 
\eeq     
Consistency with the large $L$ approximation is easily verified.
{On the other hand}, given $L$, the critical value ${\phi_c}$ increases with $u_L$ from the minimal value ${\phi_c \sim \pi}$ 
to the limiting value ${\phi_c \sim (L/2)\pi}$. 
   
More generally, the conclusion is that a condensate ${n_k=N\delta_{k,k_m}}$ of a rotating ring is energetically stable if ${|\phi-\phi_m|<\phi_c}$. This means that the potential floor ${V(\phi, n_k)}$ features parallel {\em grooves} in the $\phi$ direction. It follows that the number of flow-states for a given $\phi$ equals ${\sim \phi_c/\pi}$, and increases with $u_L$ from~1 to~$L/2$.

\subsection{Meissner Grooves}

In addition to the short-range interaction $U$ there is also a long-range electrostatic interaction $\mathcal{U}_{EM}$. We shall discuss the implication of this interaction, on equal footing with the $U$ interaction, in \Sec{sMeiss}. The bottom line, based on Bogoliubov analysis of the low excitations, is that the Grooves are further deepened and in some sense an excitation gap can form.   

Concluding this section: {\em the interaction has two opposing effects}. On the one hand it creates Grooves that can support meta-stable condensates for fixed $\phi$, which is the explanation of Landau for the feasibility of super-flow. On the other hand, it couples the different washboard-sections, leading to ergodization, and hence worsening the conditions for meta-stability. Eventually, for large enough interaction, superflow is diminished due to the formation of Mott insulator.

\section{The SF-SC border}
\label{sBorder}

For small $\alpha$, the only metastable states that can exist are SF-type as implied by the Landau stability condition. But as $\alpha$ is increased, the question arises whether there are {\em additional} valleys at ${\phi_m \ne 0}$ that can support SC-type metastable states. We refer to this as the \mbox{SF-SC} transition. The existence of SC-type metastable states depends on the model parameters, notably on $\alpha$ and $u$, as  discussed here and in subsequent sections. 

For weak interaction the condition for having a barrier between the ${k_0}$ condensate and the ${k_1}$ condensate can be worked out as follows:
Define $E=f_k(\phi)$ as the energy \Eq{eHosc} for a condensation in orbital ${k=0,\pm}$ given $\phi$. In \Fig{fBO} the green curve $f_0(\phi)$ and the blue curve $f_{+}(\phi)$ frame two Born-Oppenheimer washboard-sections that intersect at ${\phi=\pi}$. This is where we can have a barrier between the two valleys that accommodate condensates. The condition for having this barrier is ${f'_{+}(\phi=\pi)<0}$, leading to the identification of the SC regime: 
\beq \label{eSFSC}
2\pi\frac{\omega_0}{\alpha}  \ < \ 
\frac{1}{L}\sin\left(\frac{\pi}{L}\right) \ NK
\ \ \ \
\eeq
%
%
Note that the two sides correspond to $W_{\Phi}$ and $W_K$ respectively. The ratio between them corresponds to the dimensional parameter $w_L$ whose significance has been discussed in \Sec{sFloor}.

If non-zero interaction $U$ is introduced, the transitions that were neglected in \Eq{eHn}, lead to some depletion, say of $\sim n_{dep}$ particles, and consequently the effective value of $N$ in \Eq{eSFSC} becomes smaller. For uniformly distributed depletion we get  
${ N_s = N-[L/(L{-}1)]n_{dep}}$. 
In the extreme case, for equal occupation of the orbitals, as in the MI regime, we get ${N_s=0}$. The implication is that for strong interaction (large $U$) the meta-stability diminishes.

\section{Characterization of eigenstates} 
\label{sMeasures}

In the previous sections we have discussed the topography of the energy floor, and the implication on the existence of metastable flow-states. But we would like to adopt a more comprehensive perspective that allows identification of regions in the spectrum,  semiclassical-related classification of the eigenstates, and detection of fingerprints of underlying chaos. For this purpose we define several useful measures.          

Given $(N, N_{osc}, K, U, \omega_0, \alpha)$ we diagonalize the Hamiltonian \Eq{eH}, get the eigen-energies $E_{\nu}$ and the eigenfunctions $\Psi^{(\nu)}_{n_{ring},n_{osc}}$. Here the index ${n_{ring} \equiv \{ n_k \} }$ indicates the Fock orbital-occupation basis states, and $n_{osc}$ indicates the occupation basis states of the EM mode. For a trimer ring the dimensionality of Hilbert space is  
${\mathcal{N}=\mathcal{N}_{trimer}\mathcal{N}_{osc}}$, 
where ${\mathcal{N}_{trimer}=(1/2)(N{+}1)(N{+}2)}$ 
and ${\mathcal{N}_{osc}=(N_{osc}{+}1)}$.

\subsection{Occupation measures} 

For each eigenstate we calculate the the orbital occupations $\braket{\bm{n}_k}$. For the trimer we label the three orbitals as ${k=0,\pm}$. In the figure we use RGB color-code as follows: 
\beq
RGB = \left(\frac{\braket{\bm{n}_{-}}}{N},\frac{\braket{\bm{n}_{0}}}{N},\frac{\braket{\bm{n}_{+}}}{N}\right)
\eeq
If we focus on the EM mode, then complementary information 
is provided by $\braket{\bm{\phi}}$ and $\braket{\bm{n}_{osc}}$.

\subsection{Condensation measures} 

The one-body reduced probability matrix of particles in the ring,  
and the reduced probability matrix of the oscillator, are 
\beq
\rho^{(ring)}_{k,k'} &=& \frac{1}{N} \braket{a_k^{\dagger} a_{k'}} 
\\
\rho^{(osc)}_{n.n'} &=& \sum_{n_{ring}} \Psi_{n_{ring},n}  \Psi_{n_{ring},n'}^* 
\eeq 
There is no disorder in our Hamiltonian, and therefore $\rho^{(ring)}_{k,k'}$ is diagonal.    
Consequently, we define condensation and entanglement measures: 
\beq
\mathcal{S}_{purity} &=& \text{trace}\left([\rho^{(ring)}]^2\right)  
= \sum_n  \left[ \frac{\braket{\bm{n}_k}}{N} \right]^2
\\
\mathcal{S}_{ent} &=&  \text{trace}\left([\rho^{(osc)}]^2\right)  
\eeq
Roughly, $1/\mathcal{S}_{purity}$ is the number of participating orbitals, such that $\mathcal{S}_{purity}=1$ implies condensation in a single orbital. Irrespective of that, $\mathcal{S}_{ent}=1$ implies that the ring-osc eigenstate can be factorized, which is the case for either full SF condensation or strict MI fragmentation.   We define ${\mathcal{N}_{ent} = 1/\mathcal{S}_{ent} }$, and note that ${\log[\mathcal{N}_{ent}]}$ is like entanglement entropy.

\subsection{Ergodicity measures} 

The participation number $\mathcal{M}$ tells us how many basis Fock states participate in the formation of an eigenstate. Considering either site or orbital basis we extract the probabilities ${p_{n_{ring}} = \sum_{n_{osc}}|\Psi_{n_{ring},n_{osc}}|^2}$. Then we calculate the participation number   
\beq
\mathcal{M} = \left[\sum_{n} p_n^2 \right]^{-1}
\eeq
See e.g. \cite{bem}, and note that this is equivalent to the calculation of generalized fractal dimension as in e.g. \cite{BHHchainChaos1,BHHchainChaos2}. An individual eigenstate is possibly not ergodic, and does not fully accommodate the energetically allowed space. In order to determine the volume of the allowed space, we calculate the averaged $p_n$ within the energy window of interest, and then calculate the associated participation number which we denote as $\overline{\mathcal{M}}$.  
The ratio $\mathcal{M}/\overline{\mathcal{M}}$ serves as a quantum ergodicity measure. For a fully chaotic system such as billiard one expects it to be somewhat less than unity due to fluctuations.  In practice the value is much smaller indicating lack of ergodicity.   
We calculate both $\mathcal{M}_{\text{sites}}$ in the site basis, and  $\mathcal{M}_{\text{orbitals}}$ in the orbital (momentum) basis, and plot   
\beq
\mathcal{M} \equiv \min\left[\mathcal{M}_{\text{sites}},\mathcal{M}_{\text{orbitals}} \right]
\eeq
Few words are in order regarding the reasoning behind this definition. Let us discuss a general (abstract) system with basis states $\ket{n}$. Consider an eigenstate $\ket{E}$ and associated probability matrix 
${\rho_{n,m} = \Braket{n}{E}\Braket{E}{m} }$. More generally consider $\rho_{n,m}$ that is obtained by averaging over a small energy window. Clearly ${ \tr[\rho^2] }$ reflects the number of eigenstates that have been mixed.  Furthermore, we have the inequality  ${ \tr[\rho^2] > \sum_n p_n^2 \equiv \mathcal{S}_{basis} }$, where ${p_{n}=\rho_{n,n}}$. Formally we can say that  ${ \tr[\rho^2] }$  is the maximum  ${ \mathcal{S}_{basis} }$ over all possible bases. Or equivalently we say that ${\mathcal{M}}$ is the minimum participation over all bases. If we restrict the minimization over semiclassically-meaningful bases, we get an operative definition for the volume of the {\em energy shall}.  If the individual eigenstates are say localized, then the semiclassical volume ${\overline{\mathcal{M}}}$ of the energy shell becomes relatively large, with negligible dependence on the energy-width of the averaging window, see e.g. demonstration in \cite{bht}.  Accordingly the ratio $\mathcal{M}/\overline{\mathcal{M}}$ provides a practical estimate for the fraction of the energy-shell that is occupied by an individual eigenstate.

\section{Tomography of the spectrum}
\label{sTomog}




We demonstrate numerically spectrum tomography for the simplest circuit, namely, a trimer ring ($L{=}3$). Depending on parameters such ring can support either an SF ground-state or SC multi-stability. {If we take \Eq{eAlpha} literally, the realistic value of $\alpha$ is very small. Considering an ${\mathcal{L}\sim 10\mu \text{m}}$ ring it comes out ${\alpha \sim 10^{-4}}$ for an optical cavity, and much smaller for Microwave cavity. In a numerical demonstration one is limited by ${N \sim 50}$, and therefore, to illustrate an SF-SC transition, we artificially compensate by setting unrealistic large~$\alpha$. In any case, one should be aware that to regard the model as an actual representation of a realistic experimental system is rather challenging.}   

Note that the Landau criterion implies that SF meta-stability (as opposed to SC meta-stability) requires more then four sites (${L>4}$). With 3~sites, SF meta-stability can be demonstrated if the ring is rotating (with ${\phi \sim\pi}$), see Appendix~H of \cite{bem}, which in the present context would be equivalent to the introduction of an external flux $\Phi_{ext}$ as explained in \App{sSQUID}.

\begin{figure*}

\includegraphics[width=5cm]{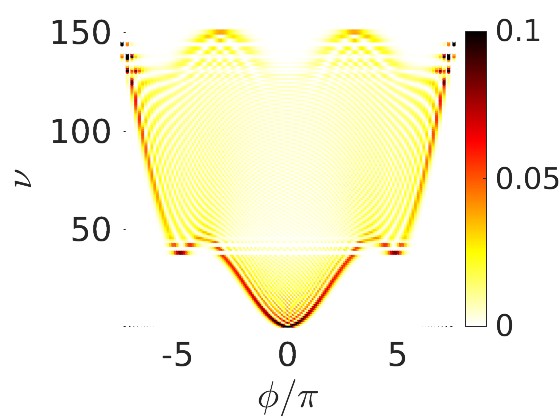}
\includegraphics[width=5cm]{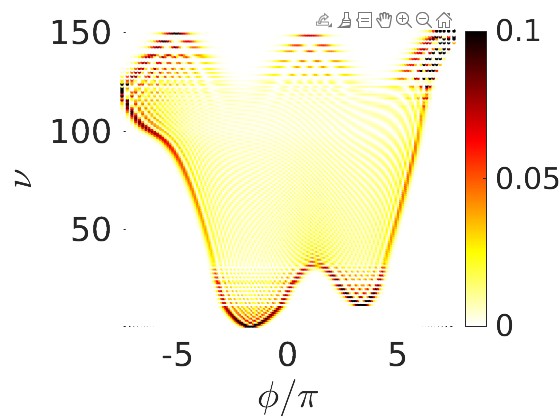}
\includegraphics[width=5cm]{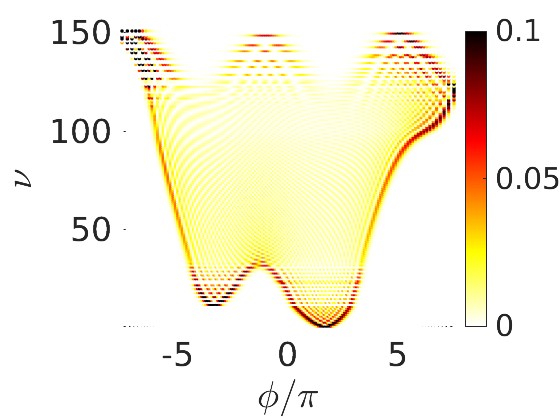} 

\includegraphics[width=5cm]{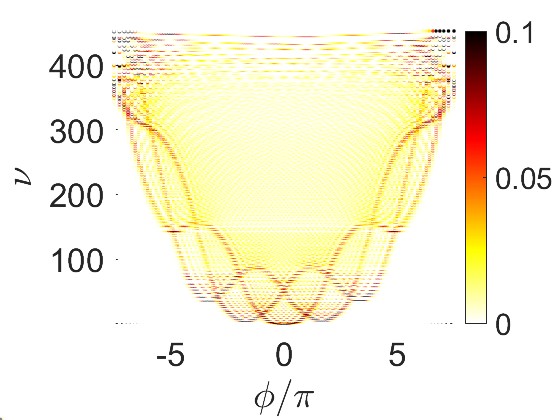}
\includegraphics[width=5cm]{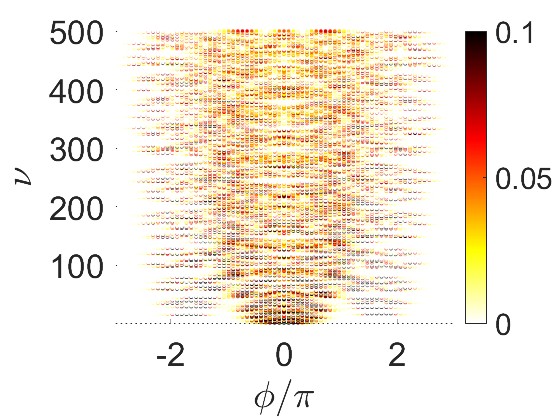}
\includegraphics[width=5cm]{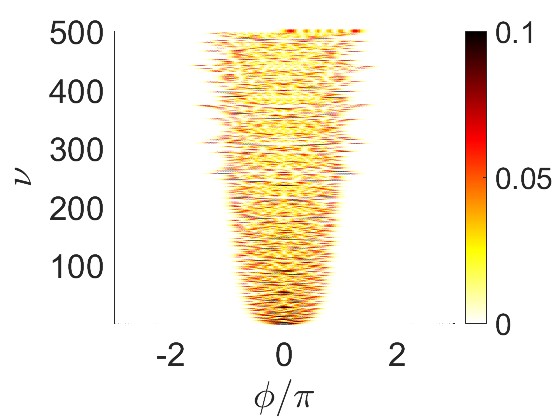}

\caption{
{\bf The energy landscape:}
We consider $u{=}0$ trimer with ${N=12}$ particles coupled to an EM mode. 
The other parameters are ${\omega_0 = 0.48}$ and ${\alpha=2}$, while ${N_{osc}=150}$. 
The upper panels provide images for eigenstates of $\mathcal{H}_{osc}$ \Eq{eHosc}   
assuming that the particles are condensed in one of the three $k$ orbitals. 
{Each row of a panel is an image of} the $\phi$ probability distribution. 
{The rows within each panel} are ordered by energy. 
The lower left panel combines the 3~upper upper panels. 
It would be the full spectrum if we had a single particle.  
For ${N=12}$ one has to combine the spectra that are associated with all possible occupations. The outcome (zoomed) is the middle lower panel.
In the right lower panel we display the result for ${\alpha=1/2}$, 
where SC multi-stability is absent. 
}  
\label{fEnLandscape}  
%
%
\ \\  \ \\ \ \\ 
%
%

\includegraphics[width=5cm]{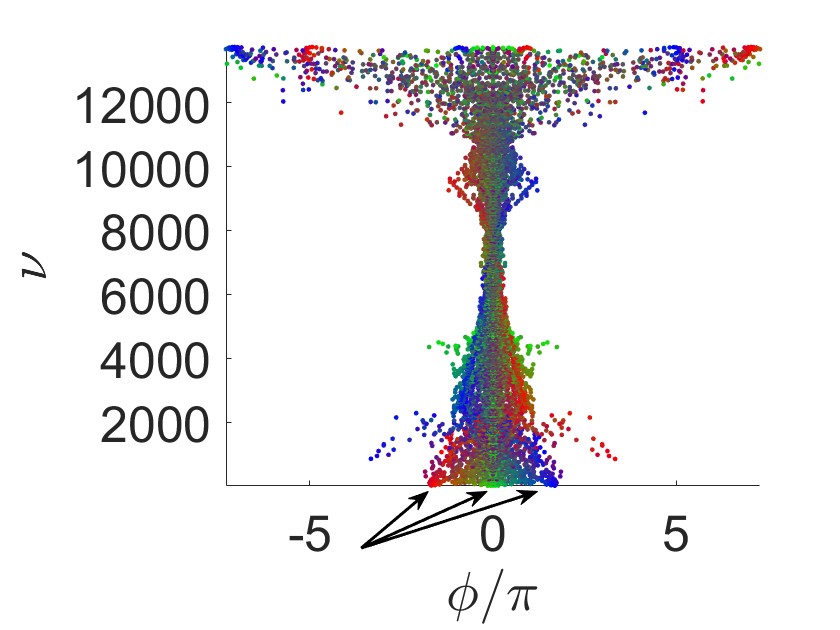}
\includegraphics[width=5cm]{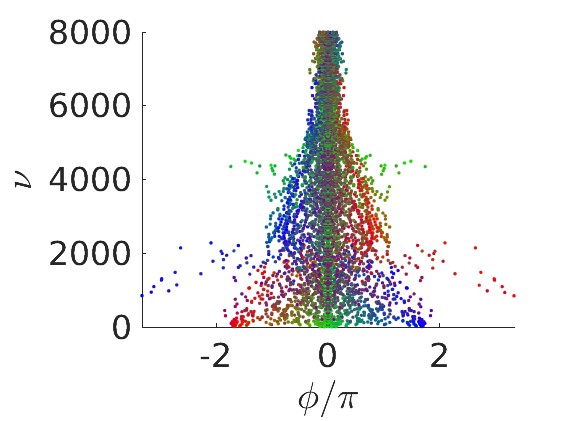}
\includegraphics[width=5cm]{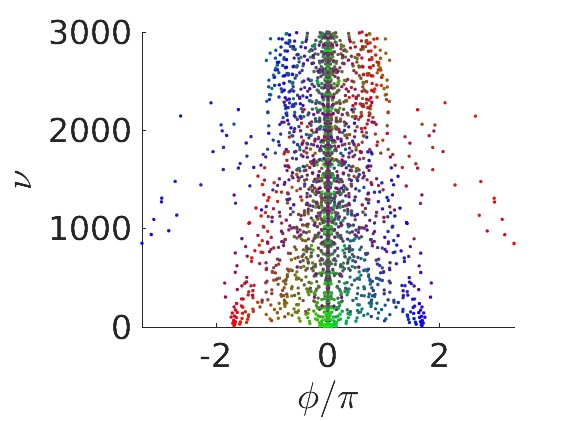}

\includegraphics[width=5cm]{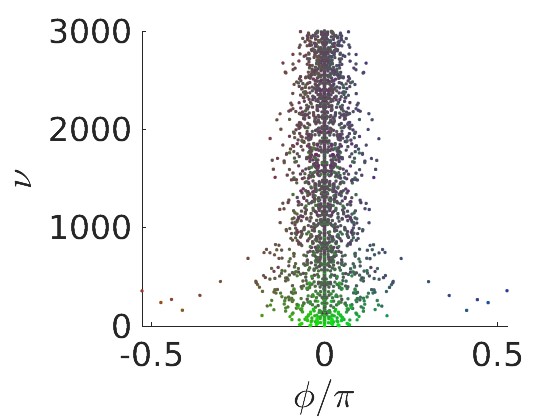}
\includegraphics[width=5cm]{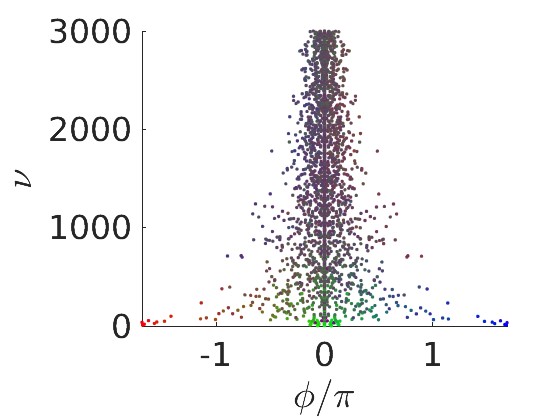}
\includegraphics[width=5cm]{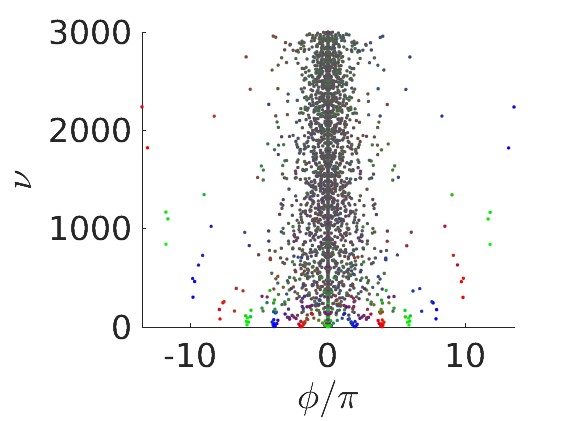}

\caption{
{\bf Tomography of the spectrum:}
The spectrum for trimer with $N=12$ particles, 
coupled to an EM~mode with $\omega_0=0.48$ and $N_{osc}=150$. 
Each eigenstate is represented by a point that is positioned according to its energy index $\nu$ and its $\braket{\phi}$. The RGB color-code indicates the occupation of the orbitals.  
In the first row the interaction is ${u=1/2}$ and the coupling is ${\alpha=2}$.
{In the left-most panel the arrows are pointing towards the green ground-state and towards the major red and blue meta-stable states}. The two right panels are zoomed versions of the left panel, {showing with better resolution this SC multi-stability, and excluding the upper false eigenstates that appear due to finite $N_{osc}$.}
In the second row the interaction is ${u=3}$ and the couplings, from left to right, are ${\alpha=1/2,2,20}$. The panels are zoomed over the first $3000$ states. This set demonstrates the crossover from SF to SC multi-stability. Note that the range of the horizontal axis has been adjusted.      
}  
\label{fTomog}  
\end{figure*}

\begin{figure*}

{\bf (a)} \hspace{6cm} {\bf (c)}

\includegraphics[width=7cm]{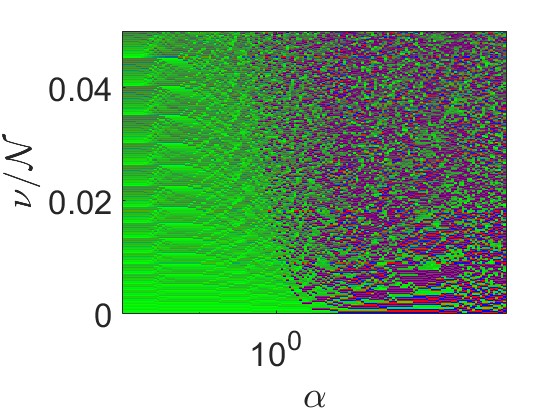}
\includegraphics[width=7cm]{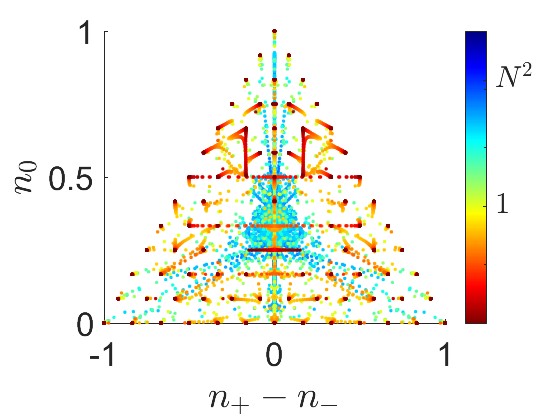}

{\bf (b1)} \hspace{5cm} {\bf (b2)} \hspace{5cm} {\bf (b3)}

\includegraphics[width=5cm]{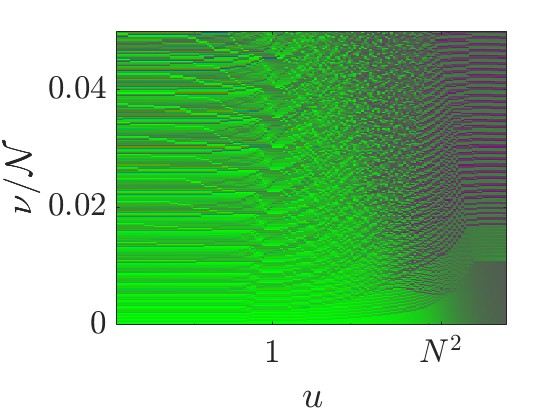}
\includegraphics[width=5cm]{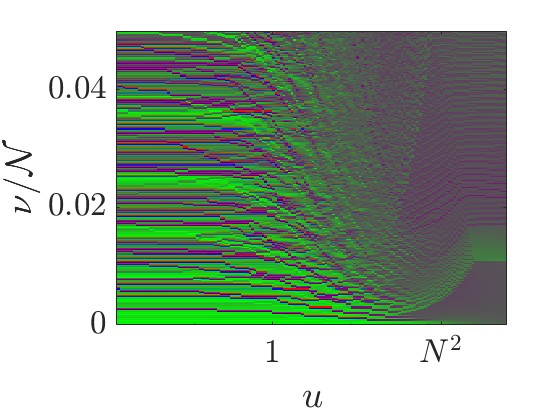}
\includegraphics[width=5cm]{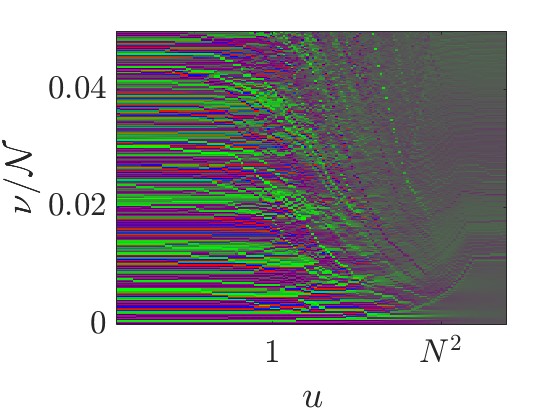}

\caption{
{\bf The formation of SC multi-stability:}
(a)~{Each column of the panel is an image of the spectrum} for a different value of $\alpha$. The energy levels (indexed by $\nu$) are RGB color-coded, reflecting orbital occupation as explained in the text. The interaction between the particles is weak (${u=0.01}$).
The other parameters are ${\omega_0=0.48}$, and ${N=12}$ and ${N_{osc}=60}$. The states are in the range ${0<\nu<0.05\mathcal{N}}$, namely, the 277 lowest levels out of 5551 are displayed. 
%
(b)~The dependence of the spectrum on $u$ for representative values of EM coupling, namely, ${\alpha=0.1,2,20}$. This set demonstrates the crossover from SF to SC, and how SF/SC is diminished due to~$u$.   
(c) The information of panel~b3 is displayed in a different way. Each point is  positioned according to the normalized orbital occupations, and the color-code indicates~$u$. As the interaction is increased multi-stability is diminished.   
}  
\label{fMS}  
%
%
\ \\ \ \\ \ \\ 
%
%

\includegraphics[width=5cm]{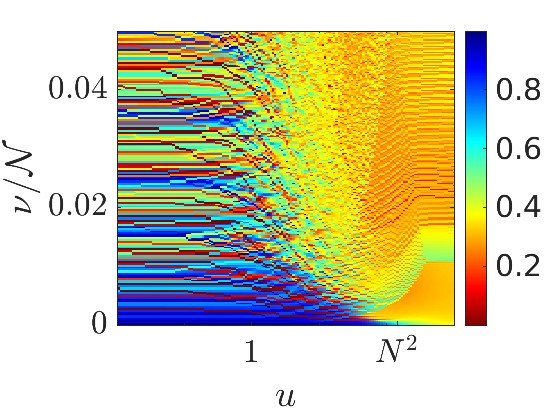} 
\includegraphics[width=5cm]{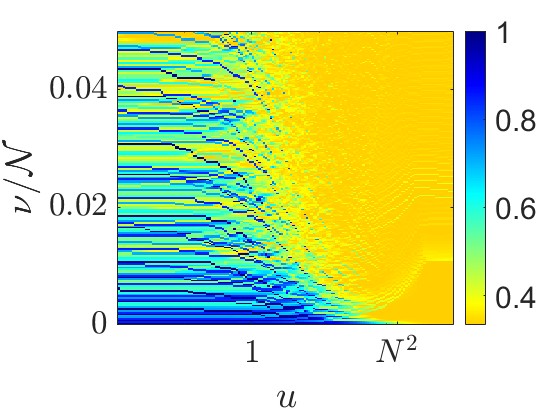} 
\includegraphics[width=5cm]{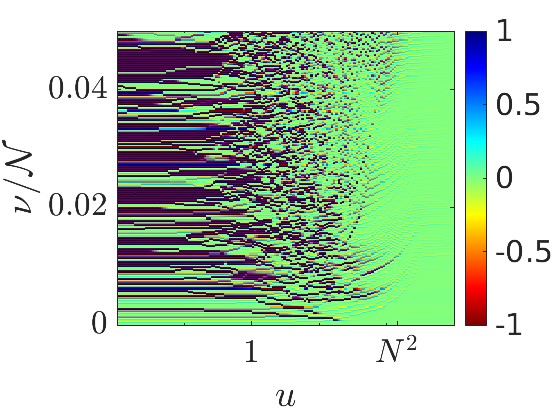}

\caption{ 
{\bf Multi-stability versus interaction.}
{Additional} diagrams that show the $u$ dependence of the spectrum. The 3~panels are: 
(a)~The normalized occupation~$n_0$ of the zero momentum orbital;
(b)~The purity $S_{purity}$ serves as a condensation measure; 
(c)~The EM mode flux $\braket{\phi}$ serves to identify SC symmetry breaking. 
The parameters are ${N=12}$, and ${N_{osc}=60}$, and ${\alpha=2}$, and ${\omega_0=0.48}$. 
}
\label{fMSvsU}  
\end{figure*}

\begin{figure*}

\includegraphics[width=5cm]{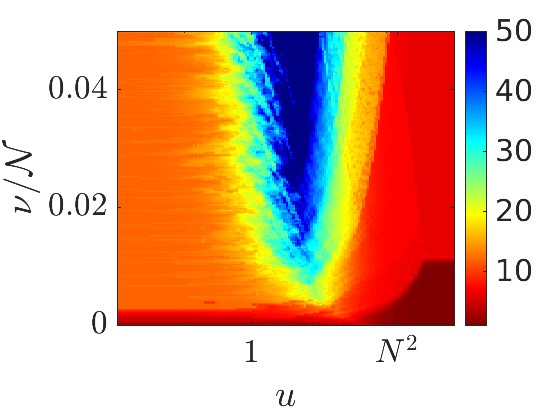} 
\includegraphics[width=5cm]{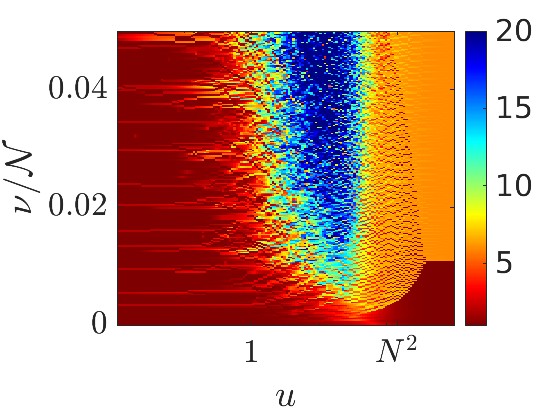}
\includegraphics[width=5cm]{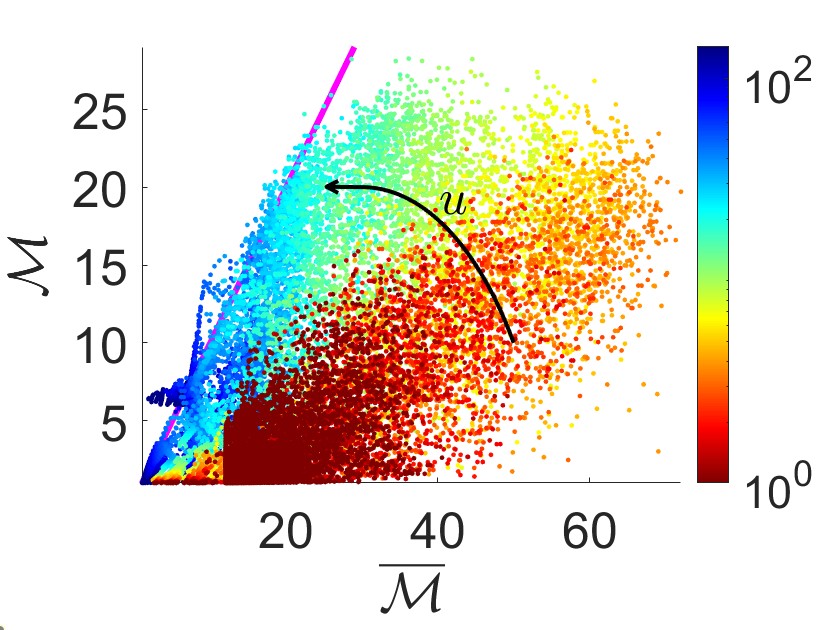}

\caption{	
{\bf {Ergodicity:}}
From left to right:  
Images of $\overline{\mathcal{M}}$ and $\mathcal{M}$. 
The parameters are the same as in \Fig{fMSvsU}.
Scatter diagrams color-coded by $u$, for inspecting the correlation between the 2~measures. For reference we plot a diagonal magenta line that indicates full ergodicity. {The trend in the ``quantum chaos regime" is indicated by an arrow that points in the direction of increasing~$u$.} 
}  
\label{fErgMeas}  
%
\ \\ 
%

\includegraphics[width=5cm]{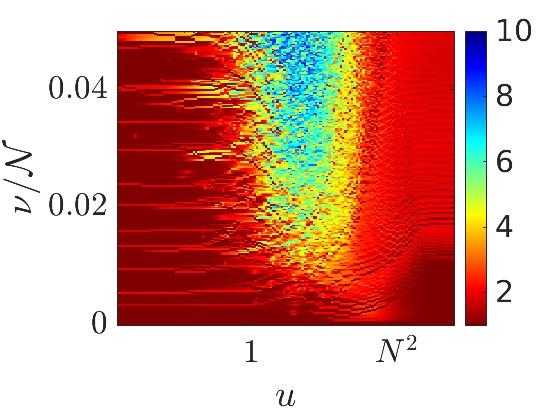} 
\includegraphics[width=5cm]{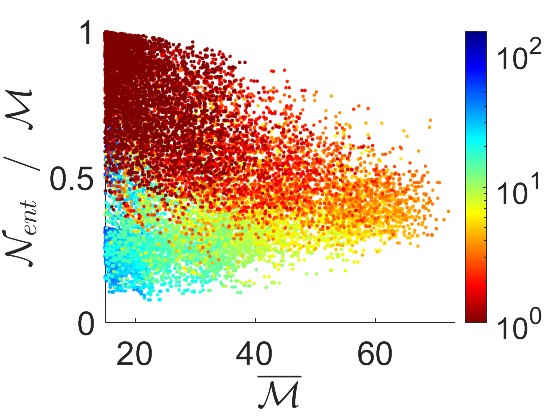}
\includegraphics[width=5cm]{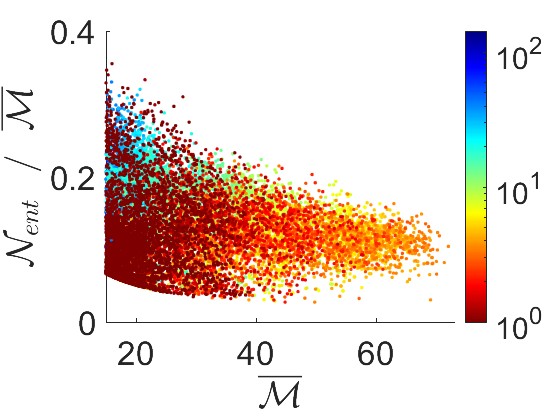}

\caption{	
{\bf {Entanglement:}}
From left to right:
Image of $\mathcal{N}_{ent}$. 
The parameters are the same as in \Fig{fMSvsU}.
{The other two panels are scatter diagrams color-coded by $u$,     
for demonstrating that $\mathcal{N}_{ent}$ is more correlated with $\overline{\mathcal{M}}$ than with~$\mathcal{M}$.}
}  
\label{fEntMeas}  
\end{figure*}

\subsection{Spectra}

\Fig{fEnLandscape} displays images for selected eigenstates of an $L=3$ circuit for $U=0$. Namely, each row is an image of the $\phi$ probability distribution for an eigenstate of \Eq{eHosc}. {The upper eigenstates are false: they appear due to the finite truncation $N_{osc}$. Namely, due to this truncation the spectrum is bounded from above}. The zoomed panels {exclude the false eigenstates.} They show images of all the low-lying eigenstates for ${\alpha=1/2}$ and  ${\alpha=2}$. The latter exhibits SC multi-stability as opposed to the former that feature an SF ground-state. 
A numerical remark is in order here. Mathematically speaking quasi-degenerated SC metastable states should combine into cat-state superpositions that feature ${\braket{\phi}=0}$, but we care to introduce a numerically negligible bias $\Phi_{ext}$ that allows `spontaneous' breaking of this symmetry.

\Fig{fTomog} displays the full spectrum for non-zero interaction. Each eigenstate is represented by a single point that is RGB color-coded such that red, green, and blue indicate the respective occupations ${(n_{-},n_0,n_{+})}$. The $n_k$ are not ``good quantum numbers" but rather are expectation values.  The selected set of panels clearly demonstrates the crossover from SF to SC multi-stability. {In all subsequent figures we care to exclude the upper false eigenstates that appear due to the finite truncation $N_{osc}$. Whatever is displayed is independent of this truncation, and the identification of the various regimes (SF-SC-FR-MI) is not affected.}

\subsection{Regimes}

We construct regime diagrams to show the dependence of the spectrum on the coupling~$\alpha$ and on the interaction~$u$. In \Fig{fMS} each column is an RGB color-coded spectrum.  
\Fig{fMS}a is an ${(\alpha,E)}$ regime diagram. Green region indicates SF states for which ${\braket{\phi}=0}$, while scrambled RGB region indicates SC multi-stability associated with broken symmetry ${\braket{\phi}\ne 0}$. The observed SF-SC transition roughly agrees with the border of \Eq{eSFSC}, and will be further discussed in \Sec{sQborders}.

\Fig{fMS}b demonstrates how the multi-stability is diminished as $u$ is increased. This is the transition to the MI regime where the eigenstates get fragmented in the Fock site basis. {Further} visualizations of this dependence in provided in \Fig{fMSvsU} where we inspect separately the occupation~$n_0$ of the zero momentum orbital; the purity $S_{purity}$ that serves as a condensation measure; and $\braket{\phi}$ that serves to identify the SC symmetry breaking.

\subsection{Ergodicity}

{The Hamiltonian is integrable in both limits of very small and very large~$u$. In particular, in the large $u$ regime, where the site-occupations become good quantum numbers, one encounters the transition to~MI. In Ref\cite{bem} we have provided a very detailed tomographic account of this transition, trying to identify eigenstates that reflect particle-hole excitations, and also amplitude and phase modes that appear at the entrance to the MI regime. It seems that for charged bosons the emerging picture is similar, we therefore do not dwell further on this theme.}    
For intermediate values of $u$ the underlying phase space is chaotic and possibly mixed with quasi-regular motion. In order to diagnose this aspect we provide in \Fig{fErgMeas} the ${(u,E)}$ diagrams for both $\overline{\mathcal{M}}$ and $\mathcal{M}$. We see that the accessible phase space region expands for intermediate values of~$u$, indicated by $\overline{\mathcal{M}}$. 

{We further construct in \Fig{fErgMeas} a scatter diagram to characterize the degree of ergodicity. The magenta line indicates the ergodic limit ${\mathcal{M}/\overline{\mathcal{M}} \sim 1}$. The arrow on top of the scatter diagram indicates that there is a trend towards `better' ergodicity in the increasing $u$ direction. The inspected region of the diagram is where $\overline{\mathcal{M}}$ is large (say more than~20). This is the region where a semiclassical perspective of ergodicity is meaningful. The small value of the $\mathcal{M}/\overline{\mathcal{M}}$ ratio in the MI region is misleading because it is outside of this semiclassical region. Following a more detailed inspection in \cite{bem}, the reason for getting `better' ergodicity for large $u$ is the shrinking of $\overline{\mathcal{M}}$ (accessible phase-space), while throughout the $\mathcal{M}$ of individual eigenstates is small and much less affected.}

\subsection{Entanglement}

We provide in \Fig{fEntMeas} the ${(u,E)}$ diagrams for the entanglement measure $\mathcal{N}_{ent}$. The associated scatter diagrams show that it is correlated with $\overline{\mathcal{M}}$ and not much with $\mathcal{M}$. In other words, the entanglement reflects the geometry of the accessible energy surface and not its chaoticity.  Irrespective of that $\mathcal{N}_{ent}$ is bounded by $\mathcal{M}$, which is responsible to a residual breakdown of the correlation in the small $u$ region. 

To explain the correlation with $\overline{\mathcal{M}}$, we point out that by definition $\mathcal{N}_{ent}$ is a measure for the influence of $n_{osc}$ on the accessible phase-space region of the ring. Maximum entanglement is achieved if the relative regions are `orthogonal'. Whether the motion in a given accessible region is ergodic or not, does not affect much the overlap between the relative regions.

\begin{figure}

\includegraphics[width=8cm]{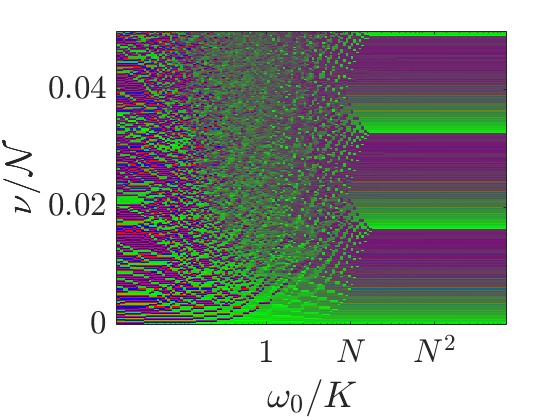}	
\includegraphics[width=8cm]{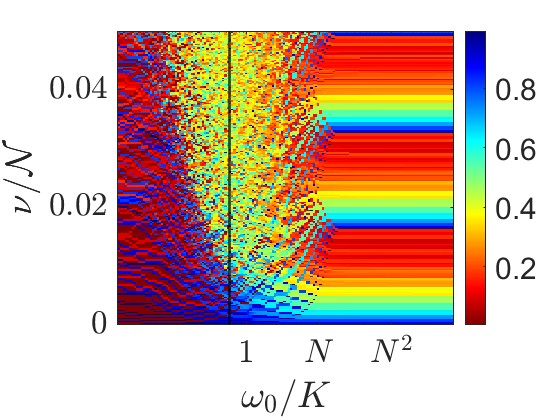}	
\includegraphics[width=8cm]{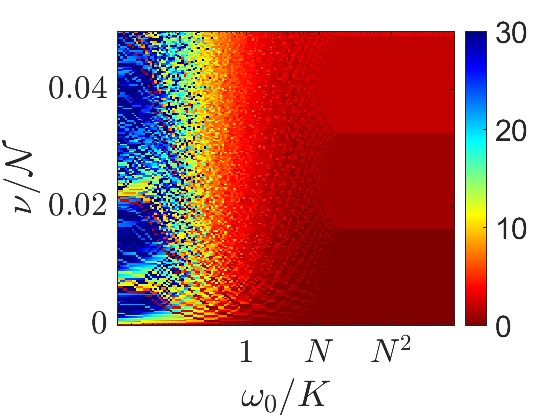}

\caption{	
{\bf Multi-stability versus mode frequency.}
The diagrams show the dependence of the color-coded spectrum on the mode frequency $\omega_0$. 
The 3~panels from top to bottom are:
(a)~RGB-coded occupation;
(b)~The normalized occupation~$n_0$ of the zero momentum orbital;
(c)~The occupation $n_{osc}$ of the cavity mode.  
The other parameters are $\alpha=2$ and $u=3$, while $N=12$ and $N_{osc}=60$. 
As $\omega_0$ is increased the low states go through  
an SC region (where we have multi-stability); 
and an SF regime (where we have single-stability).
Above the SC-SF region we have a fragmented regime (FR).    
To the right of the SC-SF-FR regimes we have an $n_0$ rainbow region where $n_{osc}=0,1,2$.  
}
\label{fMSvMF}  
\end{figure}

\begin{figure}[b!]

\includegraphics[width=4cm]{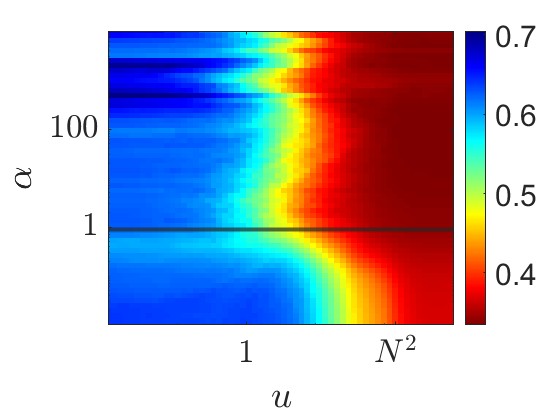}
\includegraphics[width=4cm]{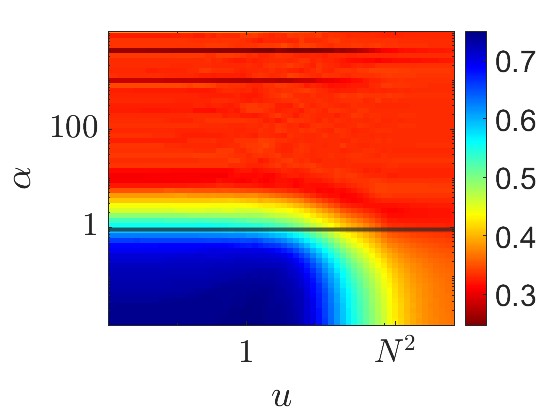}	

\includegraphics[width=8cm]{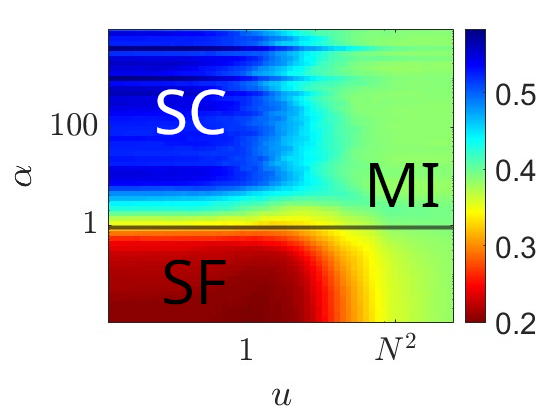} 

\caption{{\bf SF-SC-MI regime diagram.}
The transitions between these regimes is detected by looking on the dependence of averaged measures on model parameter ${(u,\alpha)}$. Other parameters are $N=12$ and $N_{osc}=60$ and ${\omega_0=0.48}$. The average is over the eigenstates in the energy range ${0<\nu<0.05\mathcal{N}}$. 
The 3~panels are:
(Left-)~The measure $\overline{\mathcal{S}_{purity}}$ indicates condensation; 
(Right-)~The measure $\overline{(n_0/N)}$ indicates zero orbital condensation; 
(Bottom-)~An ad-hoc combination of the previous panels, 
namely, ${[1{-}\overline{(n_0/N)}](\overline{\mathcal{S}_{purity}})^{1/2}}$, 
effectively provides the desired regime diagram.
The horizontal black line is the expected SF-SC metastability border.
}  
\label{fRD}  
\end{figure}

\subsection{Borders}

The dependence of the spectrum 
on $\omega_0$ is shown in \Fig{fMSvMF}, 
featuring SC-SF-FR regions.   
It is further discussed in the next section.

Placing the focus only on the low-lying levels, 
one can construct an SF-SC-MI diagram 
as demonstrated in \Fig{fRD}. 
This diagram highlights the $\alpha$-related SF-SC border of \Eq{eSFSC}, and the $u$-related MI transition that is further discussed in the next section.  In order to identify numerically the 3~regimes we have improvised an ad-hoc measure that is based on $\mathcal{S}_{purity}$ and on the zero-momentum orbital occupation~$n_0$.

\section{The SC-SF-FR spectral regions}
\label{sQborders}

The tomography allows the identification of SC-SF-FR-MI regions in the full quantum spectrum. As model parameters are varied, it is not just the ground state that is affected. We already identified in previous section the MI region using ${(u,E)}$ diagrams. We now focus on the identification of the SC-SF-FR regions, keeping away from the MI region (i.e. we assume small $u$). The most illuminating figure is possibly \Fig{fMSvMF}b with complementary panels~(a,~c). In order to understand the illustrated dependence of the spectrum on $\omega_0$ we have to clearly distinguish between classical and quantum borders. The classical borders are determined by the classically-scaled parameters ${(NU/K, \ \omega_0/K, \ N\alpha)}$, whereas quantum borders involve extra $N$ dependence ($1/N$ plays the role of Planck constant). For example the MI border is ${u_L \sim N^2}$ for the ground state and ${u_L \sim N}$ for excited states as discussed in \cite{bem}.          

The semiclassical analysis of meta-stability has implicitly assumed that the topography of the energy floor, and in particular of the valleys and the grooves, can be resolved quantum mechanically. The quantum borders are implied by inspection of the conditions for that. From the perspective of \Eq{eHosc}, the question is whether or how many eigenstates 
can be accommodated by a given valley. 

As a preliminary stage let us discuss the quantum significance of $\alpha$. If ${\alpha < 1}$ there is a sequence of ${\sim 1/\alpha}$ lowest states that have energy spacing $\sim \omega_0$, and are contained in the $k_0$ valley. The next states belong to the $k_1$ valley whose energy floor is higher, namely $W_{\Phi}=\omega_0/\alpha$. From now on let us assume for simplicity that ${\alpha > 1}$, meaning that subsequent states belong to different valleys. Below we call it the ``scrambled SC region" of the spectrum.

{In the tomographic image of the spectrum we can count how many eigenstates we have in each valley if the RGB color-code is used. In a scrambled region e.g. panel~(b3) of \Fig{fMS}, we see a mix of the 3~colors. In \Fig{fMSvMF}b the color-code discriminates the $k_0$ condensate (blue) from the $k_{\pm}$ condensates (red) because it reflects the occupation of the zero-momentum orbital. Accordingly, the scrambled  region features mix of blue and red eigenstates, indicating multi-stability.}

The question arises, how many eigenstates can be accommodated by each valley. The rough estimate is determined by the inequality  
\beq \label{eCount}
\nu \ < \ \frac{W_{K}}{\omega_0}  
\eeq
The border of the scrambled SC region in \Fig{fMSvMF}b 
is determined by \Eq{eCount}. More precisely it should be multiplied by the number of valleys that are being filled (3~in our numerical example).  
The $1/\omega_0$ dependence of its border is apparent.

We continue the discussion of regimes that appear in \Fig{fMSvMF}b. 
As $\omega_0$ is further increased the classical SC-SF border is crossed, and then only the blue eigenstates remain. Above the SC-SF region we see yellowish eigenstates, indicating the FR region. The color-code indicates that the occupation is fragmented as opposed to the condensation that is implied or hinted by either blue or red color respectively.          

{Finally, let us discuss the right-most quantum `rainbow' region that we see in \Fig{fMSvMF}b. This is the region where $\omega_0$ is so large, such that the oscillator is `frozen'. Within the lowest rainbow, all the eigenstates feature ${\nu_{osc}=0}$ (vacuum state), and zero-point fluctuations around the minimum (${\phi \sim n_{osc} \sim 0}$). The color-code merely distinguishes different disentangled eigenstates of a ring that experiences ${\phi\sim 0}$. The blue color of the lowest eigenstate in the rainbow reflects condensation in the zero momentum orbital.} Assuming negligible~$u$, the orbitals form an energy band of width~$2K$, and therefore all the possible manybody excitations of the ring are in the range whose width is~$2NK$, {(or somewhat larger due to the interaction term)}. Accordingly, the rainbow regime appears in the range  ${\omega_0 > NK}$.  Consequently, the spectrum {\em separates} into sequences of states that have the same $\nu_{osc}$, as confirmed by \Fig{fMSvMF}c.

\section{Meissner effect}
\label{sMeiss}

The Meissner effect in the simplest way is explained as a screening effect duo to the coupling of the ring to the EM mode. See \App{sSQUID} for a concise summary that clarifies that the dimensionless screening parameter is $w_L$ of \Eq{ewL}. An optional way to express this dimensionless parameter is ${w_L = (\omega_p/\omega_0)^2 }$, where the effective plasma frequency is 
\beq \label{wp}
\omega_p^2  \ = \ \alpha \omega_0  \frac{N K}{L^2}  
\ \sim \ \frac{e^2}{M} \left( \frac{N}{\mathcal{R}^3} \right)  
\eeq
In a more careful treatment, as presented below, we have to distinguish between $\omega_p$ that involves the total number~$N$ of particles, and $\omega_s$ that involves the effective number~$N_s$ of superfluid carriers.    

The Meissner effect can be regarded as a Higgs mechanism that provides mass to the EM mode. In a field-theory perspective the bare EM model have very small frequency $\omega_0$ that goes to zero for large cavity (small wavelengths), meaning that the EM field is massless. Due to the coupling to the condensate the EM field acquires a mass (hence the London penetration distance). This mass (frequency) is proportional to the {\em condensate density}. 
Here we have a mesoscopic version of the Higgs mechanism. Namely, the bare frequency $\omega_0$ is increased.
In the Harmonic approximation, based on \Eq{eHn}, it becomes    
\beq \label{eMeiss}
\omega = \sqrt{\omega_0^2 + \omega_s^2}   
\eeq
where $\omega_s^2$ is defined as $(1/2) \omega_p^2$ of \Eq{wp} with $N$ replaced by $N_s$.

Acquiring extra mass is not the whole story of the Anderson-Higgs mechanism \cite{PWA1,PWA2,PWA3,TDGL}. For bulk superconductor, the associated statement is that the Goldstone excitations become gapped and can be regarded as an extra longitudinal mode of the EM field. This statement requires further clarification and adaptation in the mesoscopic context. The gap is actually created due to the electrostatic interaction $\mathcal{U}_{ES}$ of \Eq{uES}. Namely, any long wavelength modulation of the charge-density has a large energy cost $\sim \omega_p$.  

Let us see how the electrostatic interaction looks like for our Bose-Hubbard ring. We can express $\mathcal{U}_{ES}$ as a function of of the site occupation operators. For this purpose we define  
\beq
\delta n_j \ = \ \bm{a}_j^{\dagger} \bm{a}_j - \frac{N}{L} 
\ = \ \frac{1}{L} \sum_{k' \ne k} a_{k'}^{\dagger} a_{k} e^{-i\mathscr{q} r_j}  
\eeq
where $a_{k'}^{\dagger} a_{k}$ destroys particle in orbital $k$ and creates it in orbital $k'$, while ${\mathscr{q}=k'{-}k}$ is the associated change in the momentum. The position of the site along the chain is $r_j$ modulo $\mathcal{L}$, while its actual position in space will be denoted as $\bm{r}_j$. We use here units of length such that the distance between sites is unity. The electrostatic energy is 
\beq
&& \mathcal{U}_{ES} \ \ =  \ \ \frac{1}{2}\sum_{i,j} \delta n_i \delta n_j \frac{e^2}{|\bm{r}_i-\bm{r}_j|}
\\ \label{Uaa}
&& \ = \frac{e^2}{L} \sum_{\mathscr{q}} f(\mathscr{q}) 
\left[ 
a_{k}^{\dagger} \, a_{k{+}\mathscr{q}}  a_{k{+}\mathscr{q}}^{\dagger} a_{k}  
+ a_{k{-}\mathscr{q}}^{\dagger} \, a_{k}  a_{k{+}\mathscr{q}}^{\dagger} a_{k}  
\right]
\ \ \ \ 
\eeq
where 
\beq
f(\mathscr{q}) =  \frac{1}{2} \sum_{r \ne 0 } \frac{\cos(\mathscr{q} r)}{|\bm{r}|} 
\eeq
For bulk superconductor ${f(\mathscr{q}) \sim 1/\mathscr{q}^2}$ but for our one-dimensional chain ${f(\mathscr{q}) \sim \ln(1/\mathscr{q}) }$. Assuming that most of the particles are condensed in the $k{=}0$ orbital, we can use the usual Bogoliubov approximation for the interaction-induced transitions between orbitals:  
\beq
\sum_{\mathscr{q}} \Delta_\mathscr{q} \left[  a_{-\mathscr{q}}^{\dagger} a_{\mathscr{q}}^{\dagger} +  a_{\mathscr{q}}  a_{-\mathscr{q}} \right] 
\eeq
where 
\beq
\Delta_\mathscr{q} \ \ = \ \ \frac{N}{L^2}  \left[\frac{e^2}{\ell}\right] f(\mathscr{q}) 
\eeq
Here we restored in the square brackets general physical units by introducing the lattice constant~$\ell$.  

The ring Hamiltonian contains the kinetic term 
${\sum_k (\varepsilon_k+\Delta_k) n_k }$, 
based on the second term of \Eq{eHn} and the first term of \Eq{Uaa}. 
Consequently, following the standard textbook derivation, the low lying Bogoliubov frequencies are 
\beq \label{eBogo}
\omega_\mathscr{q} \ = \ \sqrt{\left[ \varepsilon_\mathscr{q} + 2\left( \frac{NU}{L} + \Delta_\mathscr{q}  \right)\right] \varepsilon_\mathscr{q}  }
\eeq
where the $NU/L$ term originates from the last term of \Eq{eHn}. 
The bottom line is that the usual BH interaction~$NU/L$ is increased due to the electrostatic interaction $\Delta_\mathscr{q}$. 

Let us discuss the implications of \Eq{eBogo}. In the absence of electrostatic interaction we get due to~$U$ the usual phononic dispersion ${\omega_\mathscr{q} = c_s \mathscr{q}}$, where ${c_s = \sqrt{NUK/L}}$ is the speed of sound. This is the gapless Goldstone mode. 
To understand what is the fate of this mode we consider first a bulk superconductor for which the $\mathscr{q}$ dependence of ${\Delta_\mathscr{q} \propto 1/\mathscr{q}^2}$ 
is canceled by the  $\mathscr{q}$ dependence of ${\varepsilon_\mathscr{q} \propto \mathscr{q}^2}$. Due to this cancellation we get a gapped mode with 
${ \omega_\mathscr{q}  \sim  [\mathcal{R}^3/(\mathcal{L}^2\mathcal{\ell})] \omega_p  }$.
But for our one-dimensional ring ${\Delta_\mathscr{q} \propto \ln(1/\mathscr{q})}$, and therefore there is no such cancellation, and the phononic dispersion ${\omega_\mathscr{q} = c_s \mathscr{q}}$ is maintained. Still,  ${c_s}$ might become very large, leading to a gap due to a finite-size effect
\beq
\omega_{min}  \ = \ c_s \frac{2\pi}{L} 
\ \sim \ \left(\frac{1}{L}\right)
\left[\frac{\mathcal{R}^3}{\mathcal{L}^2\mathcal{\ell}}\right] \omega_p 
\eeq
This finite-size gap, compared with the effective plasma frequency, contains an enhancement factor that reflects the ratio between the volume of the cavity and the ``volume" of the ring, and a $1/L$ suppression factor that reflects that the ``gap" is due to a finite-size effect.    

\section{Summary and discussion} 
\label{sF}

We recall the quote of Einstein: ``A scientific theory should be as simple as possible, but no simpler". This perspective becomes of greater significance in the mesoscopic context. In this spirit, we were motivated to focus on a minimal model that combines SF and SC and and FR and MI regimes. The SF-SC transition is controlled by the generalized fine-structure-constant~$\alpha$, while the transition to the MI phase is controlled by the interaction~$u$.    

Formally, in the absence of an EM mode, $\Phi$ can be interpreted as the scaled rotation velocity (Sagnac phase) of the ring. Experiments with such configurations have been realized in cold atom atomtronic experiments \cite{atomtronics}. The {alternative} interpretation of $\Phi$ as a dynamical variable that represents an EM mode, requires further attention \cite{Stern}. Condensation of {\em charged} bosons is not a trivial matter \cite{cbosons,BECmeiss}. The idea actually predates BCS theory \cite{preBCS}, and might be relevant to fermionic systems that feature a BCS-BEC crossover \cite{BCStBEC}.

From an intellectual perspective, a model of charged bosons on a lattice constitutes a numerically accessible minimal model that allows exploration of fundamental questions regarding the relation between fully quantum and quasi-classical treatment of devices. 
We want to better understand the implication of treating a dynamical variable (here $\Phi$) as a quantum degree of freedom rather than as a classical control parameter. Some aspects of this problem have been discussed in \cite{nfb} regarding quantum mechanically driven adiabatic passage, but the present context looks more interesting in view of related quantum irreversibility study where $\Phi$ is the control parameter \cite{toa}.          

We have presented a semiclassical tomographic approach to study the SF-SC-FR-MI regions in the ${(U,\alpha,\omega_0,E)}$ regime diagram of the circuit.  Roughly speaking there are two dimensionless {\em classical} parameters $u_L$ and $w_L$ that control the emergence of SF and CD respectively, and an additional dimensionless  {\em quantum} parameter $\gamma_L$ that controls the MI transition.  Illustrations that demonstrate the different spectral regions were provided, notably \Fig{fMSvMF}b and \Fig{fRD}c.  
The observed spectral regions reflect the underlying valley structure of the energy landscape. The latter features valleys and grooves that can support meta-stable condensates.   

Due to the lattice nature of the BH ring, there is a vast crossover-region that mediates the MI transition, where quantum-chaos is pronounced, see \Fig{fErgMeas}. We find that the EM-BH entanglement is rather correlated with the geometry of the energy surface, while ergodic measures probe the mixed-chaotic nature of the dynamics.    

Finally we have addressed the Anderson-Higgs perspective of the Meissner effect, in the present mesoscopic context, using a device-oriented approach. The extra ``mass" that is gained by the EM mode is a rather minor effect. More interesting is the electrostatic interaction that for bulk superconductor is responsible for the gap that is opened in the Goldstone excitation spectrum. For the one dimensional ring, depending on the geometry of the system, this interaction may lead to an enhanced gap due to a finite-size effect. 

\ \\ 
{\bf Acknowledgments} --  
The research has been supported by the Israel Science Foundation, grant No.518/22. 

\ \\


\appendix

\section{Cavity mode} 
\label{sCavity}

The Maxwell equation are derived from Hamiltonian 
whose conjugate coordinates are $\mathcal{E}$ and $(1/c)\mathcal{A}$. 
Namely, 
\beq
\mathcal{H} = 
\int d^3x  \left[
\frac{1}{8\pi} 
\left( \mathcal{E}^2 + (\nabla \times \mathcal{A})^2  \right) 
- \frac{1}{c} J\cdot \mathcal{A}  \right]  
\ \ \ \ 
\eeq
where $J$ is a source. This Hamiltonian decomposes in a cavity into modes that are formally like Harmonic oscillators. Keeping a single mode we can use canonical variables 
${\Phi \equiv (1/c)\Phi_{\mathcal{B}} }$ and $Q$, 
where $\Phi_{\mathcal{B}}$ is the magnetic flux through the ring, while $Q$ is a canonically conjugate coordinate that reflects the electric field. Accordingly, 
\beq
\mathcal{H} \ =  \ \mathcal{H}_{EM}(Q,\Phi) - I_{e}(t) \Phi  + \mathcal{U}_{ES}
\eeq
where $I_{e}$ is the current in the ring, and  
\beq
\mathcal{H}_{EM} \ = \ \frac{1}{2C_e}Q^2  + \frac{c^2}{2L_e}\Phi^2   
\eeq
and the electrostatic energy is 
\beq \label{uES}
\mathcal{U}_{ES} \ = \ \frac{1}{8\pi} \int  \mathcal{E}_{longitudinal}^2  \, d^3x
\eeq
The equations of motion are 
\beq
\dot{Q} \ &=& \ \frac{c^2}{L_e} \Phi - I_e(t) \\
\dot{\Phi} \ &=& \ -\frac{1}{C_e} Q
\eeq 
In the absence of a current source the free oscillations are with frequency $\omega_0$ of \Eq{eOmega} 
that reflects the linear size of the cavity. 

Let us elaborate how $L_e$ and $C_e$ are determined in practice. Disregarding the BH ring, one can choose natural coordinates $(Q_{EM},\Phi_{EM})$ that describe the EM mode such that ${C_{EM} \sim L_{EM}}$ equal the linear size of the cavity. The cavity and the ring are formally coupled circuits, with mutual inductance ${L_{\perp}\sim \mathcal{L}}$ that reflects the linear size of the embedded ring.  The circuit equation 
${ \Phi_{EM} = L_{EM} \dot{Q}_{EM} + L_{\perp} I_e }$
implies that the interaction term is 
${ -(L_{\perp}/L_{EM}) I_e \Phi_{EM}  }$. 
We identify ${ \Phi = (L_{\perp}/L_{EM}) \Phi_{EM}  }$ as the flux of the EM mode via the ring. It follows that the inductance of the EM mode is ${L_e = L_{\perp}^2/L_{EM} }$. We conclude that $L_e$ is typically much smaller than the linear size of the ring.

\section{Cavity with SQUID} 
\label{sSQUID}

The Hamiltonian for cavity that includes a driving source and a SQUID-type ring can be written as \cite{Stern} 
\beq \label{eSQUID}
\mathcal{H} \ = \ \mathcal{H}_{EM} - I_{ext}(t)\Phi + \mathcal{H}_{ring}
\eeq
The ring is a chain of Josephson junctions (weak links) in series, 
namely, 
\beq \label{eRing}
\mathcal{H}_{ring} = \frac{E_C}{2} \sum_{j=1}^{L} \bm{n}_j^2 
- E_J \sum_{j} \cos\left(\varphi_{j}{-}\varphi_{j{-}1} {-} \frac{e\Phi}{L}\right) 
\ \ \ \ 
\eeq 
analogous to \Eq{eHn}. We use the notation  
\beq
\Phi_{ext} \ = \ \frac{1}{c^2}L_e I_{ext}
\eeq
and define ${\phi \equiv  e\Phi - e\Phi_{ext}}$ 
as the flux coordinate of the ring.  
Expressing $\mathcal{H}_{EM}$ with $\phi$ as in \Eq{eHn},  
this allows us to get rid of the source term in \Eq{eSQUID}, 
and the price is 
${e\Phi \mapsto \phi +  \phi_{ext} }$ in \Eq{eRing}.
It follows that the total flux inside the ring is  
\beq
\phi_{in} \ = \ \phi_{ext} + \phi  
\eeq
For very large $L$, taking the continuum limit and using quadratic approximation for the cosines, one concludes that the ${\phi}$ coordinate experiences the potential 
\beq
V(\phi) = \frac{K_e}{2} \phi^2 + \frac{K_L}{2} (\phi+\phi_{ext} - 2\pi m)^2 
\eeq
where $m$ is integer that reflects the winding number of the phase. The first term of $V(\phi)$ originates from $\mathcal{H}_{EM}$ while the second term from $\mathcal{H}_{ring}$. The definition of the coefficients is implied, namely, ${K_e=(c/e)^2/L_e}$ and 
${K_L=E_J/L^2}$. For ${e=0}$ the minimum of $V(\phi)$ is at ${\phi=0}$, meaning that the ring does not respond to the external source (no screening). For finite $e$ we get  
\beq \label{eMthin}
\phi_{\text{in}} \ \ = \ \ \frac{\phi_{\text{ext}} }{1 + w}  
+ \frac{ 2\pi m }{1+ w^{-1}} 
\eeq
where ${w=K_L/K_e}$ is the dimensionless screening parameter. In the absence of screening ($w=0$) we get the triviality ${\phi=\phi_{ext}}$, while for very large $w$ we get flux quantization. One observes that $w$ of the SQUID is the same as the $w_L$ that has been defined in \Eq{ewL}.

\section{Meissner effect for a thick ring} 
\label{sLondon}

Consider a cylindrical ring of radius $R$ that has finite thickness~$d$. We solve the equation of the magnetic field $B(r)$ within~${R<r<(R+d)}$, assuming~${d \ll R}$. To simplify the geometry we assume that the height of the cylinder is very large. Given that the density of the superconducting carries is~$n_s$, the London penetration distance is defined as 
\beq \label{elambda}
\lambda \ \ = \ \ \left( 4\pi \frac{e^2}{M c^2} n_s \right)^{-1/2} 
\eeq 
where $M$ is the mass of the particles.  
We can regard the thick ring as composed of very thin cylindrical rings. 
The radius~$r$ ring has thickness~$dr$, and encloses a flux~$\Phi(r)$.    
Hence, by the London equation, it carries a current density 
\beq
J(r) = \left[ \left(\frac{1}{2\pi r}\right) \frac{n_s}{M} \right] \ [2\pi m - e\Phi(r)]
\eeq
By geometry the flux satisfies 
\beq
\frac{d\Phi}{dr} = (2 \pi r) \frac{e}{c}B(r)
\eeq  
while Ampere law is 
\beq
\frac{dB}{dr} = -\frac{4\pi}{c} \ eJ(r)
\eeq  
Thus we get the equation 
\beq
\frac{d^2\Phi}{dr^2} \ = \ \frac{1}{\lambda^2} \ [e\Phi(r)-2\pi m]
\eeq
The solution for $\Phi(x)$ is $2\pi m$ plus a linear combination 
of $\cosh(x/\lambda)$ and $\sinh(x/\lambda)$, where ${x=r-R}$. 
The solution for $B(x)$ is obtained by taking the derivative.
The formal boundary conditions are:
${ \Phi(0)=\Phi_{\text{in}} }$ 
and ${ (e/c)B(0) = \Phi_{\text{in}} / (\pi R^2) }$ 
and ${ (e/c)B(d) = \Phi_{\text{ext}} / (\pi R^2) }$.  
From that we can get the result: 
\beq
e\Phi_{\text{in}} = \frac{e\Phi_{\text{ext}} }{\cosh\left(\frac{d}{\lambda}\right) + \frac{R}{2\lambda}\sinh\left(\frac{d}{\lambda}\right)}  
+ \frac{ 2\pi m }{1+\frac{2\lambda}{R}\coth\left(\frac{d}{\lambda}\right) } 
\ \ \ \ \ 
\eeq
For ${d\gg\lambda}$ the first term is exponentially suppressed, and one obtains ${e\Phi_{\text{in}} \approx 2\pi m }$, which is the usual statement of the Meissner effect for a thick ring. For a thin ring ${d\ll\lambda}$ the result becomes the same as \Eq{eMthin} with 
\beq \label{ewcyclinder}
w \ = \ \frac{\mathcal{L} d}{4\pi \lambda^2} 
\ \sim \  \frac{e^2}{c^2}\frac{N}{M} \left[\frac{1}{\mathcal{L}_{\parallel}}\right]   
\eeq
where ${\mathcal{L} = 2\pi R}$ and $\mathcal{L}_{\parallel}$ is the height of the cylindrical ring. \Eq{elambda} has been used with 
${n_s = N/(\mathcal{L}_{\parallel} \mathcal{L} d ) }$. 
The $w$ of \Eq{ewcyclinder} should be compared with the $w_L$ of \Eq{ewL}. In the latter the same expression appears but with $1/\mathcal{R}$ in the square brackets. Note however that the ${\lambda \propto \mathcal{R}^{3/2}}$ in \Eq{ewL} is merely a formal notation, unlike the  
${\lambda \propto (\mathcal{L}_{\parallel} \mathcal{L} d)^{1/2}}$ of \Eq{ewcyclinder}.



\end{document}